\documentclass[english,10pt,aps,prd,a4paper,preprintnumbers,notitlepage,floatfix,nofootinbib,showpacs,superscriptaddress]{revtex4-1}
\usepackage[english]{babel}
\usepackage{amsmath}
\usepackage{graphicx}
\usepackage{color}
\usepackage{mathrsfs}   
\usepackage{amssymb}
\usepackage{hyperref}
\usepackage{dcolumn}
\usepackage{bm}
\usepackage{graphicx}
\usepackage[sort&compress]{natbib}
\usepackage[normalem]{ulem}

\newcommand {\be} {\begin{equation}}
\newcommand {\ee} {\end{equation}}

\definecolor{greenLinks}{rgb}{0, 0.6, 0} 
\definecolor{blueLinks}{rgb}{0, 0, 0.6}
\definecolor{redLinks}{rgb}{0.6, 0, 0}
\definecolor{tempText}{rgb}{0.55, 0.10,0.67}
\definecolor{eprintLinks}{rgb}{0.4, 0.4, 0.4}
\definecolor{journalLinks}{rgb}{0.6, 0, 0}

\newcommand{\MYhref}[3][redLinks]{\href{#2}{\color{#1}{#3}}}%

\def\vev#1{\left\langle #1\right\rangle}

\usepackage{float}

\def\vev#1{\left\langle #1\right\rangle}

\def\21{$\mathrm{SU(2)_L \otimes U(1)_Y}$ }
\def\31{$\mathrm{SU(3)_c \otimes U(1)_Q}$ }
\def\SM{$\mathrm{SU(3)_c \otimes SU(2)_L \otimes U(1)_Y}$ }

\def\3211{$\mathrm{SU(3) \otimes SU(2)_L \otimes U(1)_R \otimes U(1)_{B-L}}$ }
\def\321{$\mathrm{SU(3) \otimes SU(2) \otimes U(1)}$ }
\def\422{$\mathrm{SU(4) \otimes SU(2) \otimes SU(2)_R}$ }
\newcommand {\ignore}[1]{}

\newcommand{\sm}{{Standard Model }}

\def\vev#1{\left\langle #1\right\rangle}

\def\SM{$\mathrm{ SU(3)_C \otimes SU(2)_L \otimes U(1)_Y }$ }
\def\U1{$\mathrm{ U(1)_{B_3 - 3 L_\mu} }$}

\newcommand{\AddrAHEP}{%
  AHEP Group, Institut de F\'{i}sica Corpuscular --
  C.S.I.C./Universitat de Val\`{e}ncia, Parc Cient\'ific de Paterna.\\
 C/ Catedr\'atico Jos\'e Beltr\'an, 2 E-46980 Paterna (Valencia) - SPAIN}
\newcommand{\AddrTai}{%
  Department of Physics, National Taiwan University, Taipei 10617, Taiwan}
  \newcommand{\AddrTUM}{ Physik-Department T30d, Technische Universit\"{a}t M\"{u}nchen.\\
James-Franck-Strasse, 85748 Garching, Germany}

\begin{document}

\title{\U1 gauge symmetry as a simple description of $b\to s$ anomalies}

\author{Cesar Bonilla}\email{cesar.bonilla@tum.de}
\affiliation{\AddrTUM}
\author{Tanmoy Modak}\email{tanmoyy@hep1.phys.ntu.edu.tw}
\affiliation{\AddrTai} 
\author{Rahul Srivastava}\email{rahulsri@ific.uv.es}
\affiliation{\AddrAHEP}
\author{Jos\'{e} W. F. Valle}\email{valle@ific.uv.es}
\affiliation{\AddrAHEP}


\begin{abstract}
   \vspace{1cm}
   
We present a simple \U1 gauge Standard Model extension that can easily account for the anomalies in $R(K)$ and $R(K^*)$ 
reported by LHCb. The model is economical in its setup and particle content. Among the Standard Model fermions, only 
the third generation quark family and the second generation leptons transform non-trivially under the new \U1 symmetry. This leads 
to lepton non-universality and flavor changing neutral 
currents involving the second and third quark families.  We discuss the relevant experimental constraints and some implications.

\end{abstract}

\maketitle


\section{INTRODUCTION}
\label{ref:sec1}


A host of increasingly sophisticated experiments over several decades has been able to thoroughly verify various predictions of 
the \sm of particle physics.
These culminated with the discovery of the $125$ GeV Higgs boson at the Large Hadron Collider (LHC) in 2012.
Despite its amazing success, there are good reasons to think that the \sm may not be the ultimate theory. Apart from many 
theoretical shortcomings, the existence of neutrino mass suggests the existence of new physics, possibly in the electroweak-TeV range.
By making very precise measurements of decay rates and angular observables, the LHCb Collaboration looks for the effect of new 
particles in various hadronic processes.
Of particular interest to LHCb are the processes which are either forbidden or are extremely rare within the Standard Model. 
Since such processes may be allowed in new physics models, these searches can probe new physics models with good sensitivity, 
sometimes higher than attainable at the ATLAS and CMS experiments.

LHCb has recently announced anomalous measurements of $b \to s \mu^+ \mu^-$ transitions~\cite{Aaij:2017vbb}.
They measured the ratio
$R_{K^*} \equiv {\cal B}(B^0 \to K^{*0} \mu^+ \mu^-)/{\cal B}(B^0 \to
K^{*0} e^+ e^-)$ as
\begin{equation}
  \label{eq:lhcbkb}
R_{K^*}^{\rm expt} =
\left\{
\begin{array}{cc}
0.660^{+0.110}_{-0.070}~{\rm (stat)} \pm 0.024~{\rm (syst)} ~,~~ & 0.045 \le q^2 \le 1.1 ~{\rm GeV}^2 ~, \\
0.685^{+0.113}_{-0.069}~{\rm (stat)} \pm 0.047~{\rm (syst)} ~,~~ & 1.1 \le q^2 \le 6.0 ~{\rm GeV}^2 ~.
\end{array}
\right.
\end{equation}
These measurements involve two ranges of $q^2$, the dilepton invariant
squared mass. These numbers are very similar to the previous LHCb
measurement of the ratio
$R_K \equiv {\cal B}(B^+ \to K^+ \mu^+ \mu^-)/{\cal B}(B^+ \to K^+ e^+
e^-)$ \cite{PhysRevLett.113.151601},
\begin{equation}
  \label{eq:lhcbk}
R_K^{\rm expt} = 0.745^{+0.090}_{-0.074}~{\rm (stat)} \pm 0.036~{\rm (syst)} ~,~~ 1 \le q^2 \le 6.0 ~{\rm GeV}^2 ~,
\end{equation}

These observations are also in tune with the so called $P'_5$ anomaly observed in the angular 
variable $P'_5$ of $B \to K^* \mu^+ \mu^-$ decays~\cite{DescotesGenon:2012zf,Aaij:2013qta,Abdesselam:2016llu, Wehle:2016yoi}. In 
addition to these, LHCb has also observed other several other anomalies all involving $b \to s$ transitions, such as
$B_s \to \phi \mu^+ \mu^-$~\cite{Aaij:2015esa}.  
Specially remarkable is the fact that, although each individual result is not specially significant, all of the anomalies 
observed in $b \to s \mu^+ \mu^-$ transitions and form a coherent global 
picture~\cite{Altmannshofer:2017yso,Capdevila:2017bsm,DAmico:2017mtc,Hiller:2017bzc,Ciuchini:2017mik,Geng:2017svp,Celis:2017doq,Hurth:2017hxg,Alok:2017sui}.

At a basic level such rare transitions may be described by the effective Hamiltonian,
\begin{equation}
\label{Heff}
 \mathcal{H}_{eff}=-\frac{4G_{F}}{\sqrt{2}}
 \frac{e^2}{16\pi^2}V_{tb}V_{ts}^\ast\sum_{i} \left(\mathcal{C}_{i}(\Lambda)O_{i}(\Lambda)+
 \mathcal{C}^\prime_{i}(\Lambda)O'_{i}(\Lambda)\right)
\end{equation}
where $\mathcal{C}_{i}^{(\prime)}=C_i^{(\prime)SM}+C_i^{(\prime)NP}$. Here each coefficient is separated into a \sm 
(SM) part and a new physics contribution (NP).
The relevant semileptonic operators required to account for the observed $b \to s \mu^+ \mu^-$ anomalies are of the restricted type~\cite{Descotes-Genon:2013wba},
\begin{eqnarray}
\begin{array}{cc}
 \mathcal{O}_9=(\bar{s}\gamma_\alpha P_L b)(\bar{\ell}\gamma^\alpha \ell), & 
 \mathcal{O}_9^\prime=(\bar{s}\gamma_\alpha P_R b)(\bar{\ell}\gamma^\alpha \ell) \\
 \mathcal{O}_{10}=(\bar{s}\gamma_\alpha P_L b)(\bar{\ell}\gamma^\alpha \gamma_5 \ell), & 
 \mathcal{O}_{10}^\prime=(\bar{s}\gamma_\alpha P_R b)(\bar{\ell}\gamma^\alpha \gamma_5 \ell)\notag~.
 \end{array}
\end{eqnarray}

In this paper we propose a consistent gauge model, constructed from first principles, that induces just one of the Wilson operators, $\mathcal{O}_9$. Its strength parameter can describe the observed $b \to s \mu^+ \mu^-$ anomalies in agreement with all existing experimental restrictions.
It provides a minimal way to account for the $b \to s \mu^+ \mu^-$ discrepancies, while adding as few new particles and symmetries as possible : just a new \U1 gauge symmetry.
In contrast to other alternative schemes, which typically require several the addition of new charged fermions\footnote{ For $Z'$ models without need of additional vector like quarks, see Ref. \cite{Crivellin:2015lwa, Bhatia:2017tgo}.} \cite{Altmannshofer:2014cfa, Altmannshofer:2016jzy, Crivellin:2015era,Sierra:2015fma,Carmona:2015ena, Ko:2017quv,Gauld:2013qja, Kamenik:2017tnu, Gauld:2013qba,Crivellin:2015mga,Hou:2017ozb,Fuyuto:2015gmk,Allanach:2015gkd,Celis:2016ayl,Boucenna:2016wpr,Boucenna:2016qad,Falkowski:2015zwa,Celis:2015ara,Belanger:2015nma,Chiang:2016qov,Megias:2016bde,Hou:2018npi} or leptoquarks~\cite{Alonso:2015sja,Becirevic:2016yqi,Fajfer:2015ycq,Becirevic:2017jtw,Gripaios:2014tna,Sahoo:2015wya,Cai:2017wry,Becirevic:2017jtw, Crivellin:2017zlb,Chen:2017hir,Chen:2016dip,Hiller:2016kry}, all the anomalies induced by our \U1 gauge symmetry, including the gravitational ones, cancel without need for adding any new charged states.
The only new fermions present are a pair of heavy vector-like quarks $Q'_{L,R}$ and right handed neutrinos $\nu_R$ required in order to generate small neutrino masses through the type-I seesaw mechanism~\cite{Valle:2015pba}. The only other new particles required are the $Z'$ boson associated to the new gauge symmetry and new scalars involved in symmetry breaking, so as to generate mass for the $Z'$ boson and neutrinos. The $Z'$ boson mediates the anomalous $b \to s \mu^+ \mu^-$ transitions.

The plan of the paper is as follows. In Section \ref{ref:sec2} we discuss the basics of the \U1 gauge model including the gauge and gauge-gravity anomaly cancellation conditions. We show that the model is free from anomalies. 
In Section~\ref{sec:model} we summarize the main properties of the model and show that the flavor changing neutral currents (FCNC) mediated by the $Z'$ boson have all the essential features required in order to account for the observed $b \to s \mu^+ \mu^-$ discrepancies.
In Section~\ref{sec:exper-constr} we discuss the various constraints on our model coming from flavor, 
collider and precision physics. We show that, after taking the relevant constraints into account, the model still has 
enough freedom to account for the observed anomalies in the B-system. Finally, our results are given in Fig.~\ref{fig:allowed} 
and Sec.\ref{sec:Results} and are summarized in Sec.~\ref{sec:conclusions}.


\section{Gauging the \U1 symmetry }
\label{ref:sec2}


In this section we discuss the details of the model. The \SM and \U1 charge assignments for the fermions and 
scalars are given in Table \ref{tab1}, where $i = 1, 2$ labels the first two generations of quarks.
  \begin{table}[h]
\begin{center}
\begin{tabular}{c c c || c c c}
  \hline \hline
Fields             \ \     & $SU(3)_c \times SU(2)_L \times U(1)_Y$ \ \                 & $U(1)_{B_3 - 3 L_\mu}$    \ \     & 
Fields             \ \     & $SU(3)_c \times SU(2)_L \times U(1)_Y$ \ \                 &$U(1)_{B_3 - 3 L_\mu}$                        \\
\hline \hline
$Q_{i}$ \ \   & $(3, 2, \frac{1}{3})$ \ \   &  $0$   \ \   & $L_e$ \ \   & $(1, 2, -1)$\ \  &  $0$         \\
$u_{R_i}$ \ \   & $(3, 1, \frac{4}{3})$  \ \  &  $0$ \ \   & $e_R$  \ \  & $(1, 1, -2)$  \ \   &  $0$         \\
$d_{R_i}$ \ \   & $(3, 1, -\frac{2}{3})$ \ \  &  $0$ \ \   & $\nu_{R_e} $  \ \    & $(1, 1, 0)$ \ \   &  $x_1$    \\
$Q_{3} $ \ \    & $(3, 2, \frac{1}{3})$  \ \   & $1$\ \    & $L_\mu $   \ \    & $(1, 2, -1)$\ \   &  $-3$         \\
$t_R$ \ \    & $(3, 1, \frac{4}{3})$  \ \   &  $1$  \ \    &$\mu_R$   \ \    & $(1, 2, -1)$ \ \   &  $-3$          \\
$b_R$ \ \    & $(3, 1, -\frac{2}{3})$ \ \   &  $1$  \ \    &$\nu_{R_\mu}$ \ \    & $(1, 1, 0)$ \ \ &  $x_2$\\
$Q'_{L} $ \ \    & $(3, 2, \frac{1}{3})$  \ \   &  $1$\  \ & $L_\tau$\ \    & $(1, 2, -1)$    \ \   &  $0$  \\
$Q'_{R} $ \ \    & $(3, 2, \frac{1}{3})$  \ \   &  $1$\  \ & $\tau_R$\ \    & $(1, 1, -2)$           \ \   &  $0$           \\
  & &                                                      & $\nu_{R_\tau}$\ \    & $(1, 1, 0)$    \ \   &  $x_3$   \\
\hline
$\Phi_1$\ \    & $(1, 2, -1)$  \ \   &  $1$\ \         &$\chi$ \ \    & $(1, 1, 0)$      \ \   &  $y_1$ \\
$\Phi_2$\ \ & $(1,2, 1)$\ \ &  $0$\ \                         & $\sigma $\ \    & $(1, 1, 0)$ \ \   &  $y_2$\ \ \\
\hline \hline
  \end{tabular}
\end{center}
\caption{Particle content and  assignments. The singlet $\sigma$ ensures a realistic pattern of neutrino oscillations~\cite{Forero:2014bxa}.}
  \label{tab1}
\end{table} 
Notice that apart from the ``\sm like'' $SU(2)_L$ doublet scalar $\Phi_2$, we also have added another $SU(2)_L$ doublet scalar $\Phi_1$ which is also charged under the new \U1 symmetry. 
Notice that we have also included two \SM singlet scalars $\chi$ and $\sigma$ which are also charged under 
the \U1 symmetry. As we will discuss shortly, the scalar $\chi$ is required to ensure that the model does not have a massless 
Nambu-Goldstone boson left after symmetry breaking. The scalar $\sigma$ is needed in order to ensure a realistic pattern of neutrino 
masses and mixing that can describe oscillations. To induce the latter we also include, apart from the \sm fermions, three right handed 
neutrinos $\nu_{R_i}$ ($i = e,\mu,\tau$).
Finally, we need the vector-like quarks $Q'_{L}$ and $Q'_{R}$, transforming as $SU(2)_L$ doublets, and also carrying \U1 
charges as shown in Tab. \ref{tab1}.

In order for \U1 to be a consistent gauge symmetry it is important to ensure that the model is anomaly free. The \U1 symmetry 
can potentially induce the following triangular anomalies:
\begin{eqnarray}
[SU(3)_c]^2 \, U(1)_{X}   \, \,   & \to  &  \, \,     \sum_q (X)_{q_L}  - \sum_q (X)_{q_R}    \label{cxa}   \\
\left[SU(2)_L\right]^2   \, U(1)_{X}   \, \,   & \to  &  \, \,      \sum_l (X)_{l_L}  + 3 \, \sum_q (X)_{q_L}    \label{lxa} \\
\left[U(1)_Y\right]^2   \, U(1)_{X}    \, \,   & \to  &  \, \,      \sum_{l,q} \left[ Y^2_{l_L} \, (X)_{l_L}  + 3 \, Y^2_{q_L} \,  (X)_{q_L} \right]  
\, - \,  \sum_{l,q} \left[ Y^2_{l_R} \, (X)_{l_R}  
+3 \, Yx^2_{q_R} \,  (X)_{q_R} \right] \label{y2xa} \\
U(1)_Y \, \left[U(1)_{X}\right]^2     \, \,   & \to  &  \, \,      \sum_{l,q} \left[ Y_{l_L} \, (X)^2_{l_L}  + 3 \, Y_{q_L} \,  (X)^2_{q_L} \right]  
\, - \,  \sum_{l,q} \left[ Y_{l_R} \, (X)^2_{l_R}  + 3 \, Y_{q_R} \,  (X)^2_{q_R} \right]  \label{yx2a}
 \end{eqnarray}
 In addition we have anomaly conditions involving the \U1 just with itself and with gravity,
\begin{eqnarray}
\left[U(1)_{X}\right]^3     \, \,   & \to  &  \, \,      \sum_{l,q} \left[ (X)^3_{l_L}  + 3 \, (X)^3_{q_L} \right]  
\, - \,  \sum_{l,q} \left[ (X)^3_{l_R}  + 3 \, (X)^3_{q_R} \right]  \label{x3a}\\
\left[\rm{Gravity}\right]^2   \, \left[U(1)_{X}\right]     \, \,   & \to  &  \, \,      \sum_{l,q} \left[ (X)_{l_L}  + 3 \, (X)_{q_L} \right]  
\, - \,  \sum_{l,q} \left[ (X)_{l_R}  + 3 \, (X)_{q_R} \right]
  \label{gxa}
 \end{eqnarray}
 where $(X)_i$ denote the \U1 fermion charges. Using the assignments in Table.~\ref{tab1}, we find that the first four 
 anomalies, i.e. eqs.~(\ref{cxa})-(\ref{yx2a}), cancel, irrespective of the charges of the right handed neutrinos.  The anomaly 
 cancellation conditions eqs.~(\ref{x3a}) and (\ref{gxa}) give the following two conditions on the \U1 charge of the right handed neutrinos
\begin{eqnarray}
 x_1^3 \, + \, x_2^3 \, + \, x_3^3 & =  & -27  \\
 x_1 \, + \, x_2 \, + \, x_3 & = & -3
 \label{rnua}
\end{eqnarray}
The only solution for eq.~(\ref{rnua}) is given by
\begin{eqnarray}
 x_i = -3, \, \, x_j = - x_k.
\end{eqnarray}
This implies that under the \U1 symmetry, one of the right handed neutrinos should transform as $-3$ while the others can 
carry any arbitrary equal and opposite charge. For definiteness and keeping in mind simple scenarios of neutrino mass generation, 
we choose to assign the following charges to the right handed neutrinos:
\begin{eqnarray}
 \nu_{R_e} & = & \nu_{R_\tau} \, \sim \, 0, \quad \nu_{R_\mu} \, \sim \, -3 
\end{eqnarray}

Also, we like to remark that the vector-like nature of the additional quarks $Q'_{L,R}$ implies that the anomaly 
cancellation conditions do not fix their charges. Thus, they can have any charge under \U1 symmetry. However, for the sake of simplicity 
and keeping minimality in mind (see Goldstone boson discussion below) we set their charges to be $1$, the same as the charges of the third 
generation of quarks.

Once the scalars get vacuum expectation values (vevs) the \SM $\otimes$ \U1 symmetry is broken down to $\mathrm{U(1)_{\rm{EM}}}$. Notice that, since only $\Phi_1$, $\chi$ and $\sigma$ are charged under the \U1 symmetry, the latter is broken only by the vevs of these scalars. Notice also that both \sm doublet scalars $\Phi_1$ and $\Phi_2$
contribute to the breaking of the \SM symmetry. 

The \U1 charges of the scalars are not fixed by anomaly cancellation conditions. But given the \U1 charges of all the fermions in 
the model, the charges of scalars can also be restricted by other considerations. 
The \U1 charge of the doublet scalar $\Phi_1$ is fixed by requiring an adequate pattern of quark mixing consistent with experiments.
The charge of singlet scalar $\chi$  is fixed by the requirement that if the \U1 charge of $\chi$ is not the same as the charge 
difference between $\Phi_1$ and $\Phi_2$, then the scalar potential will have a residual global $U(1)$ symmetry leading to a massless 
Goldstone boson. This can be avoided by taking
\begin{eqnarray}
 \chi & \sim & +1
\end{eqnarray}
which provides a term in the scalar potential like $V\supset \kappa \, ( \Phi^\dagger_1 \Phi_2 \, \chi + h.c.)$, where $\kappa$ is a dimensionful parameter. 
With these assignments for $\nu_{R_i}$, $\Phi_1$, $\Phi_2$ and $\chi$ plus a scalar singlet $\sigma$, transforming as $\sigma\sim 3$, one also has a realistic pattern of
neutrino mass and mixing~\footnote{We stress that the above choice of $\nu_{R_i}$ charges under the \U1 symmetry is just the simplest one consistent with current neutrino oscillation data~\cite{Forero:2014bxa}. A detailed treatment of neutrino properties is outside the scope of this letter and will be addressed separately.}.

\section{The model}
\label{sec:model}

Having satisfied the anomaly cancellation conditions we now turn to the scalar and Yukawa sectors of the model. 
As shown in Table \ref{tab1}, except for the Standard-Model-like Higgs scalar $\Phi_2$, all other scalars 
are charged under the \U1 symmetry. Here our main focus is on explaining the anomaly, so we will skip the details 
of the scalar sector (fairly standard) in this work, and focus on the Lagrangian characterizing the Yukawa sector, 
which can be written as follows
\begin{eqnarray}
-\mathcal{L}_{Y} & = &
y^{u}_{3j} \, \bar{Q}_{3}  u_{j} \Phi_1 
\, + \, y^{u}_{4j} \, \bar{Q}'_{L}  u_{j}  \Phi_1 
\, + \, y^{u}_{ij} \, \bar{Q}_{i}  u_{j} \tilde{\Phi}_2  
\, + \, y^{u}_{33} \, \bar{Q}_{3}  u_{3} \tilde{\Phi}_2 
\, + \, y^{u}_{43} \, \bar{Q}'_{L}  u_{3} \tilde{\Phi}_2 
\nonumber \\
& + & y^{d}_{i3} \, \bar{Q}_{i} b_{R} \tilde{\Phi}_1
\, + \, y^{d}_{ij} \, \bar{Q}_{i} d_{j} \Phi_2
\, + \, y^{d}_{33} \, \bar{Q}_{3} d_{3} \Phi_2 
\, + \, y^{d}_{43} \, \bar{Q'}_{L} b_{R}\Phi_2 
\, + \, y_{14} \bar{Q}_{1} Q'_{R}\chi^* \, 
\nonumber \\
& + & y_{24} \bar{Q}_{2} Q'_{R} \chi^*
\, + \, \mu \bar{Q}_{3} Q'_{R} 
\, + \, M_{Q'} \bar{Q}'_{L} Q'_{R} + h. c. \, 
\label{yuk}
\end{eqnarray}
where $\tilde{\Phi} = i \tau_2 \Phi^*$ and $i,j = 1,2$ represent the first two families.  
After spontaneous symmetry breaking the scalars acquire vevs $\vev{ \Phi_i} = v_i$; $i = 1,2$, $\vev{ \chi} = v_\chi$ and $\vev{ \sigma} = v_\sigma$.
The resulting quark mass matrices in the basis $\left( \bar{Q}_{1}, \bar{Q}_{2}, \bar{Q}_{3}, \bar{Q}'_L \right)^T$ and $(q_{1}, q_{2}, q_{3}, Q'_R)$ are 
given by

\begin{eqnarray}
\mathcal{M}_d = 
\left( \begin{array}{cccc}
y^d_{11} v_{2}   & y^d_{12} v_{2}   &  y^d_{13} v_{1}   & y_{14} v_\chi          \\
y^d_{21} v_{2}   & y^d_{22} v_{2}   &  y^d_{23} v_{1}   & y_{24} v_\chi    \\
0   &  0                            &  y^d_{33} v_{2}   &  \mu                 \\
0   &  0                            &  y^d_{43} v_{2}      & M_{Q'}                \\
\end{array} \right)\ \ \text{and}\ \
\mathcal{M}_u =  
\left( \begin{array}{cccc}
y^u_{11} v_{2}   & y^u_{12} v_{2}   & 0      &  y_{14} v_\chi                 \\
y^u_{21} v_{2}   & y^u_{22} v_{2}   & 0      &  y_{24} v_\chi                \\
y^u_{31} v_{1}   & y^u_{32} v_{1}   & y^u_{33} v_{2}      &   \mu                  \\
y^u_{41} v_{1}   & y^u_{42} v_{1}   & y^u_{43} v_{2}      & M_{Q'}                \\
\end{array} \right)
\label{mud}
\end{eqnarray}

The resulting charged current weak interactions of quarks and leptons can be easily generated taking into 
account the mixing of \sm quarks with the new vector-like quarks.
The mass matrices of \eqref{mud} have enough freedom to be able to generate the required Cabibbo-Kobayashi-Maskawa (CKM) mixing matrix
of charged current interactions. However, the presence of vector-like quarks charged under the \U1 symmetry leads to interesting implications 
for the neutral currents.
Hence we focus here on the weak neutral current, mediated both by the \sm $Z$ boson as well as the new $Z'$.
Since the scalars $\Phi_1$, $\chi$ and $\sigma$ are all charged under \U1, their vevs break the \U1 symmetry and contribute to 
the $Z^\prime$ mass.  Thus, in the limit in which $\vev{ \chi},\vev{ \sigma} \gg v_{1}$ and taking negligible $Z-Z^\prime$ mixing, 
the neutral gauge boson masses are given by,
\begin{equation}
 m_{Z^\prime}^2\varpropto g^{\prime2} u^2 \ \ \text{and} \ \
 m_Z^2\varpropto (g_1^2 + g_2^2) v^2
\end{equation}
where $g^\prime$, $g_1$ and $g_2$ are the coupling constants of the \U1, U(1)$_Y$ and SU(2)$_L$ symmetries, respectively. The 
vevs of the doublet scalars $\vev{\Phi_i} = v_{i}$ ($i = 1,2$) satisfy $v^2\equiv (174\, \text{GeV})^2 = v_{1}^2 + v_{2}^2 $, 
whereas $u^2 \equiv v_\chi^2 + 9 v_\sigma^2 +  v^2_{1} $.\\

The key part of the theory in order to account for the B-decay anomaly is the neutral current. One can easily see 
that the \sm part of the neutral current, involving only $Z$ boson interactions, is canonical, obeying the Glashow-Iliopoulos-Maiani 
mechanism. 
In contrast however, by construction, the neutral current Lagrangian associated with the $Z'$ boson will give rise to 
flavor changing transitions, that is 
\begin{eqnarray}
\label{Zpcurrent}
 -\mathcal{L}_{Z'}= g^\prime Z^{\prime\alpha} J'^0_\alpha
 =  
 g^\prime Z^{\prime\alpha} \left[
 -3\bar{\mu}\gamma_{\alpha} \mu
 +\bar{b}\gamma_{\alpha} b
 +\bar{t}\gamma_{\alpha}  t 
 -3\bar{\nu}\gamma_{\alpha} \nu
 +\bar{Q}'\gamma_{\alpha} Q'
 \right]. 
 \label{current}
 \end{eqnarray}
 From Eq.~\eqref{current} it is clear that in our model, in the gauge basis, the $Z'$ couples vectorially to the third generation \sm quarks and the second generation leptons. A chiral variant of this feature can also be obtained as in \cite{Cline:2017lvv}.   
After spontaneous symmetry breaking the part associated to down-type quarks becomes
\begin{eqnarray}
J'^0_\alpha
\supset
\left( \begin{array}{cccc}
\bar{d} & \bar{s} & \bar{b} & \bar{Q}'
\end{array} \right) D_L 
\gamma_{\alpha} \left( \begin{array}{cccc}
0 & 0 & 0 & 0\\
0 & 0 & 0 & 0\\
0 & 0 & 1 & 0\\
0 & 0 & 0 & 1
\end{array} \right)  D_L^\dagger 
\left( \begin{array}{c}
d \\ s \\ b \\ Q'
\end{array} \right)
+
\left( \begin{array}{cccc}
\bar{d}_R & \bar{s}_R & \bar{b}_R & \bar{Q}'
\end{array} \right) D_R
\gamma_{\alpha} \left( \begin{array}{cccc}
0 & 0 & 0 & 0\\
0 & 0 & 0 & 0\\
0 & 0 & 1 & 0\\
0 & 0 & 0 & 1
\end{array} \right) D_R^\dagger
\left( \begin{array}{c}
d_R \\ s_R \\ b_R \\ Q'
\end{array} \right),\notag
 \end{eqnarray}
where $D_L$ is the rotation matrix that diagonalizes $\mathcal{M}_d^2=\mathcal{M}_d \mathcal{M}_d^\dagger$ (namely,
$\text{diag}(m_d^2,m_s^2,m_b^2,M_{Q'}^2)=D_L \mathcal{M}_d^2 D_L^\dagger $)
and $D_R=\mathbb{I}$. The squared mass matrix for down-type quarks can be taken to the form,
\begin{eqnarray}
\label{Mdsq}
 \mathcal{M}_d^2 \sim
 \begin{small}
\left( \begin{array}{cccc}
\times   &  0   & 0 & \times          \\
0   & \times   &  0   & \times    \\
0   &  0   &  \times   &  \times                 \\
\times   &  \times   &  \times  & \times  
\end{array} \right)
\end{small} 
\ \ \text{and hence}\ \
D_L\sim 
\begin{small}
\left( \begin{array}{cccc}
1   &  0   & 0 & V_{d4}          \\
0   & 1   &  0   & V_{s4}    \\
0   &  0   &  1   &  V_{b4}                 \\
V_{4d}   &  V_{4s}   &  V_{4b}  & 1  
\end{array} \right)
\end{small}
\end{eqnarray}
As a result, the interactions between $Z'$ and the down-type quarks in the mass basis can be rewritten as follows
\begin{eqnarray}
\label{Zpmassb}
-\mathcal{L}^{'}_{Z'}
\supset g' Z'_\mu~\left[
\bar{d}'_{a}\gamma^{\mu}((\Gamma_L)_{ab} P_L + (\Gamma_R)_{ab} P_R) d'_{b} \right]
\end{eqnarray}
where $d'=(d,s,b,Q')$,  
\begin{eqnarray}
\Gamma_L=\left( \begin{small}\begin{array}{cccc}
\label{GammaLR}
|V_{4d}|^2 & V_{4d}V_{4s}^* & V_{4d}V_{4b}^* & V_{4d}V_{44}^*\\
V_{4s}V_{4d}^*  & |V_{4s}|^2 & V_{4s}V_{4b}^*& V_{4s}V_{44}^* \\
V_{4b}V_{4d}^* & V_{4b}V_{4s}^*  & 1-|V_{4b}|^2 & V_{4b}V_{44}^*\\
V_{44}V_{4d}^* & V_{44}V_{4s}^*  &  V_{44} V_{4b}^* & 1- |V_{44}|^2
\end{array}\end{small} \right) \ \ \text{and}\ \
\Gamma_R=\left( \begin{array}{cccc}
0 & 0 & 0 & 0\\
0 & 0 & 0 & 0\\
0 & 0 & 1 & 0\\
0 & 0 & 0 & 1
\end{array} \right)
\end{eqnarray}
One can see that our model implies that the $R(K)$ and $R(K^*)$ get modified only by the operator 
$\mathcal{O}_9$. Hence, $C_9^{'NP}=C_{10}^{NP}=C_{10}^{'NP}=0$. The associated strength parameter is given as
 \begin{equation}
  C_9^{NP}
  = \left(\frac{8\pi^2 v^2}{e^2 \lambda_{bs}}\right)\left(\frac{3 g^{'2}  \hat{\lambda}_{bs}}{m_{Z^\prime}^2}\right)
  = -\frac{3\hat{\lambda}_{bs}}{\hat{\alpha}\lambda_{bs}} \left(\frac{g^\prime}{m_{Z^\prime}}\right)^2\notag
 \end{equation}
where $\lambda_{bs}\equiv V_{tb}V_{ts}^\ast$, 
$\hat{\lambda}_{bs}\equiv V_{4b}V_{4s}^\ast$ and $\hat{\alpha}= e^2/(8\pi^2 v^2)\approx 1.9\times 10^{-8}\,\text{GeV}^{-2}$. 
In what follows we show that our model not only qualitatively satisfies all requirements to explain the observed 
anomalies, but can also quantitatively explain them, satisfying all relevant experimental constraints.
In our numerical computations in the next section we will work in the limit 
$|\hat{\lambda}_{bs}|\simeq|\lambda_{bs}|\sim\lambda_C^2$. Also, for a concrete benchmark we take $V_{44},V_{4b}\sim1$ and $V_{4s}\sim \lambda_C^2$. 
Furthermore, in order to avoid generating unacceptable $K^0 - \bar{K}^0$ mixing we take $V_{4d}\sim0$.
In this limit we have that eq.~(\ref{GammaLR}) becomes
\begin{eqnarray}
\label{gammaLs}
\Gamma_L\sim\left( \begin{small}\begin{array}{cccc}
0 & 0 & 0 & 0\\
0  & \lambda_C^4 & \lambda_C^2& \lambda_C^2 \\
0 & \lambda_C^2  & 0 & 1\\
0 & \lambda_C^2  &  1 & 0
\end{array}\end{small} \right).
\end{eqnarray}
In this approximation the Wilson coefficient turns out to be
\begin{equation}
  C_9^{NP}   =  -\frac{3}{\hat{\alpha}} \left(\frac{g^\prime}{m_{Z^\prime}}\right)^2\label{c9eq}.
 \end{equation}

Notice that the above choice constitutes just a simple benchmark of our model. The result in the last 
equation serves to illustrate in a simple manner that 
our model provides a successful way to account for the b-anomalies from first principles.

\section{Experimental constraints}
\label{sec:exper-constr}

Having shown that the flavor changing neutral current (FCNC) mediated by the $Z'$ boson has the right form, 
we now turn to the experimental constraints on the two parameters, the $Z'$ mass and coupling strength, relevant 
for describing in our model the anomalies recently observed in rare B decays.

\subsection{B meson mixing}
\label{sec:constr-bbmixing}

The existence of a non-zero $Z'bs$ coupling induces $B_s-\bar{B}_s$ mixing at the tree-level, resulting in very 
stringent limits on $g'$ and $m_{Z'}$. Such a tree level $Z^\prime$ exchange modifies the $B_s-\bar{B_s}$ meson mixing 
amplitude $M_{12}$, which can be quantified as~\cite{Buras:2014yna}
\begin{equation}
 \frac{M_{12}}{M_{12}^{\text{SM}}}=1+\frac{g^{'2}}{m_{Z'}^2}
 \left(\frac{g^2_2}{16\pi^2 v^2}(V_{ts}V_{tb}^\ast)^2 S_0\right)^{-1}\\
\end{equation}
where $g_2$ is the $SU(2)_L$ gauge coupling and  $S_0$ is Inami-Lim function with value $\simeq2.3$~\cite{Lenz:2010gu,Buras:2012jb}. The 
mixing amplitude $M_{12}$, related to the mass difference by
$\Delta m_{B^0_s} = 2 |M_{12}|$ is measured precisely at per mill level~\cite{Olive:2016xmw}, while the calculation of $M_{12}$ suffers 
from several uncertainties. The two dominant sources of uncertainties are hadronic matrix element and CKM factor; the former is 
estimated by FLAG 2016~\cite{Aoki:2016frl} as $\sim12\%$, while the latter is $\sim 5\%$~\cite{Charles:2015gya,Bona:2006sa}. However, 
a recent accurate estimate provided by the Fermilab Lattice and MILC collaborations~\cite{Bazavov:2016nty} has improved the hadronic 
uncertainty and pushed the FLAG average down to $\sim 6$\%. Including such sources of uncertainties, CKMfitter 
constrains $\left|\frac{M_{12}}{M_{12}^{\text{SM}}}\right| < 1.32$~\cite{Charles:2015gya} while 
UTfit $\left|\frac{M_{12}}{M_{12}^{\text{SM}}}\right| < 1.28$~\cite{Bona:2006sa} at 2-standard-deviation ($2\sigma$). Note that the 
summer 2016 result~~\cite{Bona:2006sa} of UTfit includes the result of Ref.~\cite{Bazavov:2016nty}. 
In order to constrain the the parameter space of $g'$ and $m_{Z'}$, we allow new physics contribution 
in $\left|\frac{M_{12}}{M_{12}^{\text{SM}}}\right|$ up to 30\% and 15\%, which are displayed by the purple shaded region
and purple dashed line in Fig.~\ref{fig:allowed} respectively~\footnote{See Ref.~\cite{Kohda:2018xbc} for a more 
detailed discussion on the new physics contribution in $\left|\frac{M_{12}}{M_{12}^{\text{SM}}}\right|$.}.

\subsection{Neutrino trident production}
\label{sec:constr-trident}

The production of a $\mu^{+}\mu^-$ pair in the scattering of a muon neutrino in the Coulomb field of a heavy nucleus, i.e. $\nu_\mu N\to \nu_\mu N\mu^+\mu^-$, provides a sensitive probe for $g'$. In our model the correction to the trident cross section can be expressed as~\cite{Altmannshofer:2014cfa}
\begin{align}
\frac{\sigma^{\text{NP}}}{\sigma^\text{SM}}
=\frac{1+(4s_W^2+18 v^2 g^{'2}/m_{Z'}^2)^2}{1+(1+4s^2_W)^2}~\label{tri1}
\end{align}
where $v=246$~GeV and $s_W=\sin\theta_W$.  The measurement of the trident cross section by the CCFR collaboration is~\cite{Mishra:1991bv}
\begin{align}
\frac{\sigma^{\text{CCFR}}}{\sigma^\text{SM}}
=0.82\pm 0.28~\label{tri2}.
\end{align}
Utilizing Eqs.~\eqref{tri1} and allowing $2\sigma$ error in $\frac{\sigma^{\text{CCFR}}}{\sigma^\text{SM}}$, we find the excluded region shown by the blue solid line in Fig.~\ref{fig:allowed}.
%

\subsection{Lepton Flavor Universality in $Z$ boson decay}

The presence of $Z'\mu\mu$ and $Z'\nu_{\mu} \nu_{\mu}$ couplings will break lepton flavor universality (LFU) in $Z$ boson decay. 
This is manifest in the $Z$ boson couplings to muons and muon neutrinos through loop effects. The corrections to the vector and axial 
vector couplings of $Z\mu\mu$ coupling relative to the Standard-Model-like $Zee$ can be expressed as~\cite{Altmannshofer:2014cfa,Altmannshofer:2016brv}
\begin{align}
 \frac{ g_{V\mu} }{ g_{Ve} } 
\simeq \frac{ g_{A\mu} }{ g_{Ae} } 
\simeq \left| 1+ \frac{9g'^2}{(4\pi)^2} \kappa(m_{Z'})\right|;\label{mu}
\end{align}
and similarly for $Z\nu\nu$
\begin{align}
\frac{ g_{V\nu} }{ g_{Ae} } = \frac{ g_{A\nu} }{ g_{Ae} } 
\simeq \left| 1+ \frac{1}{3}\frac{9g'^2}{(4\pi)^2} \kappa(m_{Z'})\right|\label{nu}
\end{align}
where $\kappa(m_{Z'})$ is the loop factor associated with the $Z'$ loop~\cite{Haisch:2011up}, whose real part is taken to 
match the convention of Ref.~\cite{ALEPH:2005ab}. The factor 1/3 in Eq.\eqref{nu} accounts for the fact that out of three 
neutrino flavor only $Z \to \nu_\mu \bar \nu_\mu$ is affected by the $Z'$. 
The vector and axial vector couplings of $Z$ boson in Eqs.~\eqref{mu} and \eqref{nu} can be found from the average of 14 
electroweak measurements in Ref.~\cite{ALEPH:2005ab}. The relevant ones are: 
$g_{Ve} = -0.03816\pm 0.00047$, $g_{Ae} = -0.50111\pm 0.00035$, $g_{V\mu} = -0.0367\pm 0.0023$, $g_{A\mu} = -0.50120\pm 0.00054$ 
and $g_{V\nu} = g_{A\nu} = 0.5003\pm 0.0012$. We find that $g_{A\mu}/g_{Ae}= 1.00018 \pm 0.00128$ provides the most stringent 
constraint, where the uncertainties are added in quadrature. The resulting $2\sigma$ upper limit on $g'$ is shown by the red 
dashed line in Fig.~\ref{fig:allowed}.

\subsection{Constraint from $Z\to4\ell$}
\label{sec:constr-from-zto4}

The ATLAS~\cite{TheATLAScollaboration:2013nha} and CMS~\cite{Sirunyan:2017zjc} collaborations both have set upper limits on 
the branching ratio of the $Z$ boson decay to four charged leptons. The ATLAS~\cite{TheATLAScollaboration:2013nha} analysis was 
performed with Run 1 data (7 TeV + 8 TeV), while CMS~\cite{Sirunyan:2017zjc} utilized 13 TeV 35.9 fb$^{-1}$ data to set the limit 
on  $\mathcal{B}r(Z\to4\ell)$. 
In particular, the observed value reported by Ref.~\cite{Sirunyan:2017zjc} is
$\mathcal{B}r(Z\to 4\mu)=(4.83^{+0.23}_{-0.22}(\mbox{stat})^{+0.32}_{-0.29}(\mbox{syst})\pm0.08(\mbox{theo})\pm0.12(\mbox{lumi}))\times10^{-6}$,
while the SM prediction is $(4.37 \pm 0.03)\times10^{-6}$~~\cite{Sirunyan:2017zjc}.

In our model the $Z\to 4\ell$ decay will receive contributions from processes involving the $Z^\prime$ as the intermediate state, 
such as contribution from $Z\to \mu^+\mu^-Z' $ followed by $Z'\to \mu^+\mu^-$, resulting stringent bound on $g'$ for $m_{Z'}< m_{Z}$.

In order to determine the upper limit on $g'$ from $\mathcal{B}r(Z\to 4\mu)$, we utilized the \SM prediction and observed value of 
Ref.~\cite{Sirunyan:2017zjc}, with the errors in the latter symmetrized and added in quadrature. The cross sections are generated in 
the Monte Carlo event generator MadGraph5\_aMC@NLO~\cite{Alwall:2014hca}, interfaced to PYTHIA~6.4~\cite{Sjostrand:2003wg} for 
hadronization and showering and finally fed into fast detector simulator Delphes~3.3.3~\cite{deFavereau:2013fsa} so as to incorporate detector effects. 
In our analysis we adopt the PDF set NN23LO1 PDF~\cite{Ball:2013hta}. The effective Lagrangians written in
eq.~\eqref{Zpcurrent} and eq.~\eqref{Zpmassb} are implemented in FeynRules~2.0~\cite{Alloul:2013bka}. Following the analysis 
of Ref.~\cite{Sirunyan:2017zjc}, we select events with four isolated muons with two opposite
sign same flavor dimuon pairs. The muons in the quadruplet are required to be separated by $\Delta R= \sqrt{\Delta\eta^2+\Delta\phi^2}> 0.02$, 
with each of them having maximum pseudo-rapidity $\left|\eta\right| < 2.5$.  The two
leading muons in an event should have transverse momenta $p_T$ $>$ $ 20$ GeV and $ 15$ GeV respectively, while the other two muons are 
required to have $p_T > 5$ GeV. 
The four muons will constitute two same flavor oppositely charged muon pairs in an event. The pair closest to the $Z$ boson mass 
should have invariant mass $> 40$ GeV and both pairs should have invariant mass $< 120$ GeV. All pairs of oppositely charged muons must 
have invariant mass $> 4$ GeV. We finally impose an invariant
mass cut $80~\mbox{GeV} < m_{4\mu} < 100 ~\mbox{GeV}$ on the four muons in the event. Finally, demanding the SM plus NP contribution i.e. 
the total contribution from $Z'\to \mu^+\mu^-$ not to exceed the $2\sigma$ error, we overlay the solid red line ($2\sigma$ upper limit) in the 
left panel of Fig.~\ref{fig:allowed}. The limits from Ref.~\cite{TheATLAScollaboration:2013nha} are weaker than those of 
Ref.~\cite{Sirunyan:2017zjc}, which we do not show in Fig.~\ref{fig:allowed}.

\subsection{Constraint from $p p \to Z' + X \to  \mu^+ \mu^- + X$} 
\label{sec:constraint-from-p}

The $Z'$ boson will be produced at LHC predominantly by the flavor conserving $b\bar{b}\to Z'$ processes with a 
correction from flavor violating $s\bar{b}\to Z'$ (and its conjugate process). 
Hence, the search for heavy resonances in the dimuon final state by the ATLAS and CMS collaborations will constrain 
he parameter space of our model. In particular ATLAS~\cite{Aaboud:2017buh} has set a $95\%$ CL (confidence level) upper limit 
on $\sigma(pp\to Z'+ X) \mathcal{B}r(Z'\to\mu^+\mu^-)$ 
in the 150 GeV $\lesssim m_{Z'} \lesssim$ 5 TeV mass range with the 13
TeV  and 36.1 fb$^{-1}$ dataset, where $X$ conforms inclusive activity. CMS~\cite{Sirunyan:2018exx} has also
searched for heavy resonances decaying to dimuon pair in the mass
range 200 GeV $\lesssim m_{Z'} \lesssim$ 5.5 TeV with 13 TeV, also with
$\sim$ 36 fb$^{-1}$ dataset, setting a $95\%$ CL upper limit on
$R_\sigma$ defined as:
\begin{align}
R_\sigma = \frac{ \sigma(pp\to Z'+ X \to\mu^+\mu^-+X) }{ \sigma(pp\to Z + X ~ \to\mu^+\mu^-+X) }.
\end{align}
We reinterpret $R_\sigma$ and extract
$\sigma(pp\to Z' +X) \mathcal{B}r(Z'\to\mu^+\mu^-)$ by multiplying
with the \sm prediction of
$\sigma(pp\to Z + X)\mathcal{B}r(Z\to\mu^+\mu^-)=1928.0$ pb
\cite{CMS:2015nhc}.

In order to determine the upper limit, we generate matrix element (ME) of the $p p\to Z'$ process up to two additional 
jets in the final state to include inclusive contributions. The ME is then merged and matched
with parton shower (PS) following the MLM~\cite{Alwall:2007fs} matching prescription.  We restrict ourselves up to 
two additional jets due to computational limitations. It should be noted that we have not used any $K$ factor 
in our analysis.  We finally convert the observed ATLAS (CMS) $95\%$ CL upper limit 
on $\sigma(pp\to Z' +X)\mathcal{B}r(Z'\to\mu^+\mu^-)$ to constraint $g'$ and $m_{Z'}$ in the mass 
range 150 GeV $< m_{Z'} < $ 5 TeV (200 GeV $< m_{Z'} < $ 5.5 TeV) which is shown by the black (light-blue) shaded region in Fig.~\ref{fig:allowed}.

\section{Results}
\label{sec:Results}

\begin{figure}[h!]
\includegraphics[width=0.48\textwidth]{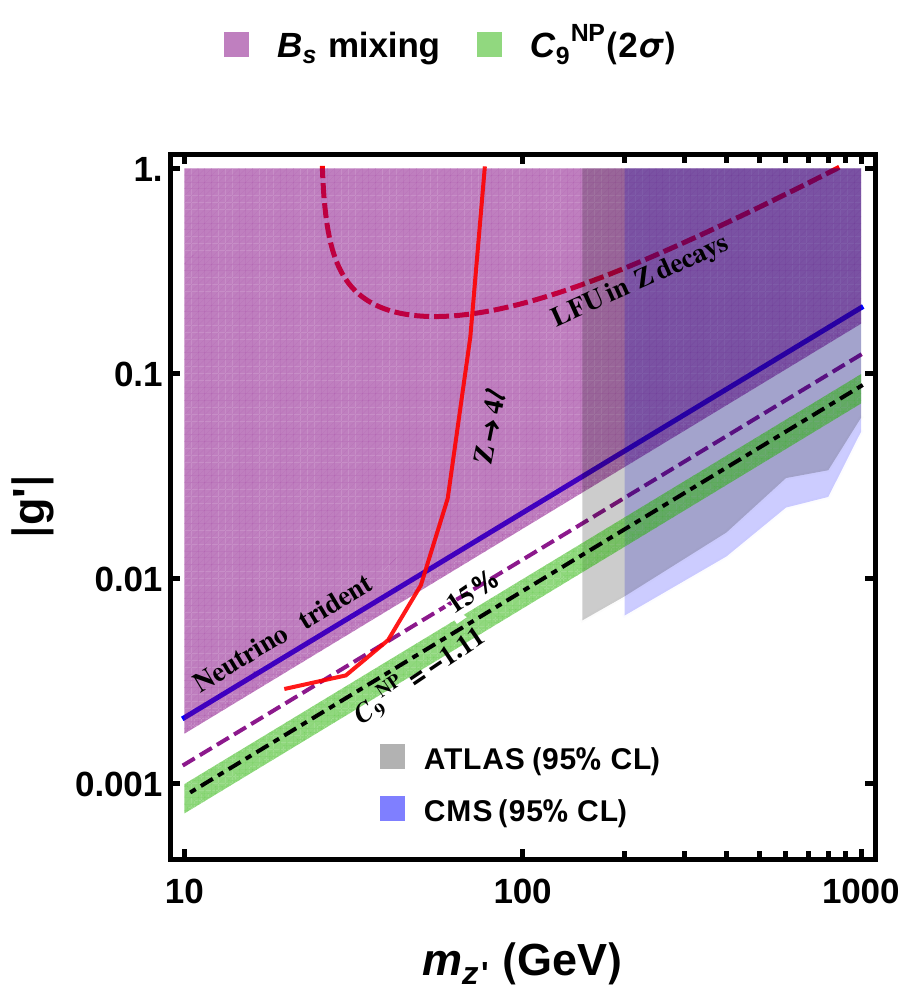}
\includegraphics[width=0.467\textwidth]{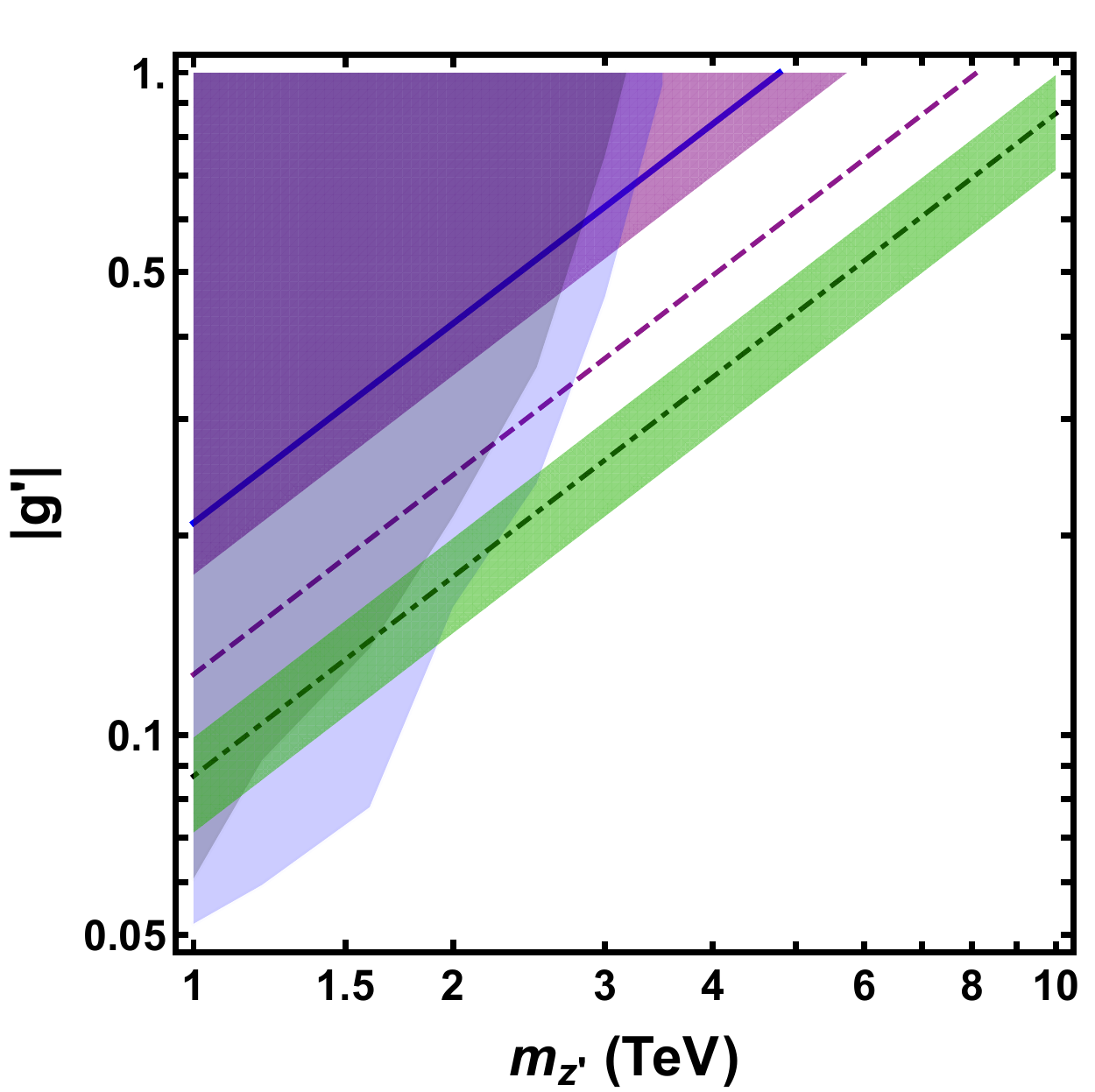}
 \caption{ Allowed region in $g'$ vs $m_{Z'}$ obtained from the simplified parameter benchmark choice made in Eqs~(\ref{Mdsq}-\ref{gammaLs}). 
See text for detailed explanation of the constraints as well as their color code description}.
\label{fig:allowed}
\end{figure}

We now discuss the constraints on the coupling $g'$ and mass $m_{Z'}$ in our model, which is summarized in Fig.~\ref{fig:allowed}.

The region of coupling-mass ($g' - m_{Z'}$) allowed by the various flavour and collider physics constraints are divided 
into two ranges indicated in Fig.~\ref{fig:allowed}. The left panel corresponds to the case of light ($m_{Z'} < 1$ TeV) masses, while 
the heavy mass range ($1 \, \rm{TeV} < m_{Z'} < 10$ TeV) $Z^\prime$ is shown in the right panel. 
In order to find the allowed parameter space which describes the observed anomalies in $b\to s \ell \ell$ transition, 
we follow the global-fit analysis presented in Ref.~\cite{Capdevila:2017bsm}.
The analysis uses all available $b\to s\ell\ell$ data from LHCb, ATLAS, CMS and Belle, with the best fit value ${\rm Re}~C^{\rm NP}_{9} = -1.11$, 
and $2\sigma$ range
\begin{align}
-1.45 \leq {\rm Re}~C_9^{\rm NP} \leq -0.75.
\label{eq:globalfit}
\end{align}
Utilizing Eq.\eqref{c9eq}, the $2\sigma$ allowed region of ${\rm Re}~C_9$ is translated into
$g'$ vs $m_{Z'}$ plane, and shown by the green shaded region in Fig.~\ref{fig:allowed},
while the central value ${\rm Re}~C^{\rm NP}_{9} = -1.11$ is shown by black dot-dashed line.
Note that the authors of Ref.~\cite{Capdevila:2017bsm} also present results taking into account only 
lepton-flavor universality observables, such as $R_{K^{(*)}}$ etc., which we do not display in 
Fig.~\ref{fig:allowed} for simplicity. 
Although in our analysis we have only utilized the global fit values from Ref.~\cite{Capdevila:2017bsm}, there exist 
other global-fit analyses based on the effective Hamiltonian formalism (see e.g. 
Refs.~\cite{Altmannshofer:2017yso,DAmico:2017mtc, Hiller:2017bzc,Geng:2017svp,Ciuchini:2017mik,Celis:2017doq,Hurth:2017hxg}), which 
can also be used with similar results as displayed in Fig.~\ref{fig:allowed}.
The resulting allowed regions are indeed very similar, with a significant overlap region~\footnote{There are other observables 
like $P'_5$, derived in a model independent way from $B\to K^* \mu^+\mu^-$ decay~\cite{Mandal:2015bsa}, which indicate new physics 
contribution in the right handed current.}.

We have also studied all relevant constraints from flavor physics as well as from LEP precision observables \cite{ALEPH:2005ab} and 
direct searches at the LHC~\cite{Aaboud:2017buh,Sirunyan:2018exx,Sirunyan:2017zjc}. The most relevant LEP constraint from these decays 
is shown by red dashed line in Fig.\ref{fig:allowed}. The region above the red dashed line is excluded by LFU in $Z$ boson decay. 
The ATLAS collaboration \cite{TheATLAScollaboration:2013nha} has also looked for the decay $Z \to 4 \mu$ which in our model is 
mediated by the $Z'$. This places a constraint on our model parameter space, shown by the solid red line (the region above the line is ruled out). 
The most stringent limits for $m_{Z'} < 150$ GeV come from $B_s$ mixing~\cite{Amhis:2014hma,Charles:2015gya}, which is shown by the purple shaded 
region in both figures; except for $ 25~\mbox{GeV} \lesssim m_{Z'} \lesssim 40$ GeV where constraint from $Z\to 4 \mu$ becomes strongest.

In the region $150~\mbox{GeV}\lesssim m_{Z'}\lesssim2.2$ TeV the constraint from $B_s$ mixing is superseded by the search for 
heavy $Z'$ boson in the dimuon final state by the ATLAS~\cite{Aaboud:2017buh} and CMS collaborations~\cite{Sirunyan:2018exx}. The black 
shaded region in Fig.\ref{fig:allowed} is excluded by ATLAS, while the light-blue shaded region is ruled out by CMS. In general, limits 
from CMS~\cite{Sirunyan:2018exx} are a bit stronger than those of ATLAS~\cite{Aaboud:2017buh}; except for 
$150~\mbox{GeV}\lesssim m_{Z'}\lesssim200$ GeV, where CMS provides no result.
One sees that the direct search limits from ATLAS and CMS rule out a simultaneous explanation of all $b \to s$ transition 
anomalies for $Z'$ masses in the range from $150~\mbox{GeV}\lesssim m_{Z'}\lesssim2.2$ TeV, leaving however the possibility for discovery 
in the range $m_{Z'} \lesssim 150$ GeV and $m_{Z'} \gtrsim 2.2$ TeV
~\footnote{See Ref.~\cite{Kohda:2018xbc} for the details on how a $Z'$, if it were confirmed in the near future, 
could be associated to $b\to s \ell \ell$ anomalies.}.
We also mention that low-energy experiments also set severe constraints on the model. For example, the constraint from neutrino 
trident production is shown by the solid blue line, with region above it excluded~\cite{Altmannshofer:2014cfa,Mishra:1991bv}. In addition, 
the decay $B \to K^{(*)} \nu\bar{\nu}$~\cite{Buras:2014fpa} can also constrain the our model parameters, though we find that the limits on $g'$ are weaker and we do not display in Fig.\ref{fig:allowed}. Finally, another potentially strong constraint can come from $D^0$-${\bar D}^0$ mixing \cite{Kumar:2018kmr}. However, we have checked that for our choice of benchmark point the limits coming from this constraint are weaker and hence not displayed in Fig.\ref{fig:allowed}.

\section{Conclusions}
\label{sec:conclusions}

In this letter we have presented a rather simple anomaly free \U1 Standard Model extension that can account for the recent 
anomalies in $b\to s \ell \ell$ transitions reported by the LHCb collaboration. 
The model is minimalistic both in its setup as well as particle content. Amongst the Standard Model quarks and leptons, 
only the third generation quark family and the second generation leptons transform non-trivially under our new postulated \U1 symmetry. 
This leads to a very simple pattern for lepton non-universality and flavor changing neutral currents involving the second and third 
families which reproduces the LHCb findings in a way consistent with all the relevant experimental constraints, except for the range 
from 150 GeV up to $\sim 2.2$ TeV, where the understanding of the $B\to K^{(*)}$ anomaly reported by LHCb would clash with the direct 
searches in the di-muon channel by the ATLAS and CMS collaborations.
One should also stress, as seen in left panel of Figs.~\ref{fig:allowed}, that the $Z^\prime$ associated to our \U1 symmetry can be 
as light as 10~GeV, in contrast to $Z^\prime$s associated to other gauge extensions based on B-L~\cite{Das:2012ii,Ma:2015mjd} or 331 
theories~\cite{Deppisch:2013cya,Queiroz:2016gif}.
As a last comment, we mention that, throughout the paper, we have assumed dominance of the vector boson mediated neutral current contribution, 
neglecting all the scalars. We checked that, indeed, there is a realistic limit in parameters space where this can be achieved.

\begin{acknowledgments}

  We thank Avelino Vicente, Masaya Kohda and Amol Dighe for useful comments and discussions. Work supported by the Spanish grants SEV-2014-0398 and FPA2017-85216-P (AEI/FEDER, UE), PROMETEO/2018/165 (Generalitat Valenciana) and the Spanish Red Consolider MultiDark FPA2017‐90566‐REDC. CB acknowledges financial support from the Collaborative Research Center SFB1258. The work of TM is supported by MOST-104-2112-M-002-017-MY2.

\end{acknowledgments}

\end{document}